\documentclass[10pt]{article}

\usepackage{fancyhdr}
\usepackage{extramarks}
\usepackage{amsmath}
\usepackage{amsthm}
\usepackage{amsfonts}
\usepackage{siunitx}
\usepackage{tikz}
\usepackage[plain]{algorithm}
\usepackage{algpseudocode}
\usepackage{multirow}
\usepackage{booktabs}
\usepackage{palatino}
\usepackage{graphicx}
\usepackage{subfigure}
\usepackage[colorlinks,linkcolor=black,anchorcolor=black,citecolor=black,urlcolor=blue]{hyperref}
\usepackage{amsmath,bm}
\usepackage{ragged2e}
\usepackage{booktabs, tabularx}
\usepackage{mathtools}
\usepackage{amssymb}
\usepackage{caption}
\usepackage{capt-of}
\usepackage{mciteplus}
\usepackage{cite}
\usepackage{mathrsfs}
\usepackage[title,titletoc,toc]{appendix}
\usepackage{xr}
\usepackage{parskip}
\usepackage{multicol}
\usepackage{multirow}
\usepackage{ragged2e}
\usetikzlibrary{automata,positioning}
\usepackage{colortbl}
\definecolor{myyellow}{rgb}{0.988,0.902,0.792}
\definecolor{mypink}{rgb}{.99,.91,.95}
\definecolor{mygreen}{rgb}{0.792,0.921,0.847}

\renewcommand{\part}[1]{\textbf{\large Part \Alph{partCounter}}\stepcounter{partCounter}\\}

\begin{document}
\title{Unveiling the molecular mechanism of SARS-CoV-2 main protease inhibition from 92 crystal structures}

\author{Duc Duy Nguyen$^1$, Kaifu Gao$^1$, Jiahui Chen$^1$, Rui Wang$^1$, and Guo-Wei Wei$^{1,2,3}$ \footnote{Address correspondences to Guo-Wei Wei. E-mail:wei@math.msu.edu} \\
$^1$ Department of Mathematics,
Michigan State University, MI 48824, USA\\
$^2$  Department of Biochemistry and Molecular Biology\\
Michigan State University, MI 48824, USA \\
$^3$ Department of Electrical and Computer Engineering \\
Michigan State University, MI 48824, USA \\
}
\date{}
\maketitle

\begin{abstract}
Currently, there is no effective antiviral drugs nor vaccine for coronavirus disease 2019 (COVID-19) caused by  acute respiratory syndrome coronavirus 2 (SARS-CoV-2).   Due to its high conservativeness and low similarity with human genes, SARS-CoV-2 main protease (M$^{\text{pro}}$) is one of the most favorable drug targets. However,  the current understanding of the molecular mechanism of  M$^{\text{pro}}$ inhibition is limited  by the lack of reliable binding affinity ranking and prediction of existing structures of M$^{\text{pro}}$-inhibitor complexes.
This work integrates mathematics and deep learning (MathDL) to provide a reliable ranking of the binding affinities of 92 SARS-CoV-2 M$^{\text{pro}}$ inhibitor structures. 	We reveal that Gly143 residue in M$^{\text{pro}}$ is the most attractive site to form  hydrogen bonds, followed by Cys145, Glu166, and His163. We also identify 45 targeted covalent bonding inhibitors. Validation on the PDBbind v2016 core set benchmark shows the MathDL has achieved the top performance with Pearson's correlation coefficient ($R_p$) being 0.858. Most importantly,  MathDL is validated on a carefully curated SARS-CoV-2 inhibitor dataset with the averaged $R_p$ as high as 0.751, which endows the reliability of the present  binding affinity prediction. The present binding affinity ranking, interaction analysis, and fragment decomposition offer a foundation for future drug discovery efforts.

\end{abstract}

\newpage


\clearpage \pagebreak \setcounter{page}{1}
\renewcommand{\thepage}{{\arabic{page}}}

\section{Introduction}
Starting in late Dec, 2019, the COVID-19 pandemic caused by new severe acute respiratory syndrome coronavirus (SARS-CoV-2) has infected more than 5.7 million individuals and has caused more than 353,000 fatalities in all of the continents and over $213$ countries and territories by May 27th, 2020. To date, there is no specific drug nor vaccine against COVID-19. Under the current global health emergency, researchers around the world have engaged in the investigation of the different drug targets of SARS-CoV-2, such as the main protease (M$^{\text{pro}}$, also called 3CL$^{\text{pro}}$), papain-Like protease (PL$^{\text{pro}}$), RNA-dependent RNA polymerase (RdRp), $5^{\prime}\text{-to-}3^{\prime}$ helicase protein (Nsp13) to seek potential cures for this serious pandemic.

The main protease, one of the best-characterized targets for coronaviruses, attracts lots of research attention because  it is very conservative and distinguished from any human gene.  A recent study shows that although the overall sequence identity between SARS-CoV and SARS-CoV-2 is just 80\%, the M$^{\text{pro}}$ of SARS-CoV-2 shares 96.08\% sequence identity to that of  SARS-CoV   \cite{xu2020evolution}. Therefore, we hypothesize that a potent SARS M$^{\text{pro}}$ inhibitor is also a potent SARA-CoV-2 M$^{\text{pro}}$ inhibitor.

At this moment, more than {300} potential SARS-CoV M$^{\text{pro}}$ inhibitors with its binding affinities are available in ChEMBL database \cite{davies2015chembl} which can be considered as the potential SARS-CoV-2 M$^{\text{pro}}$ inhibitors. Recently, total {94} crystal structures of SARS-CoV-2 M$^{\text{pro}}$ with its ligand complexes are released on the Protein Data Bank (PDB) \cite{Berman:2000}. Among them,  {92} crystal structures have no available binding affinities reported for various reasons. However,  the central dogma of drug design and discovery concerns the molecular mechanism and  binding affinity of drug target interactions. Knowing the binding affinities and their ranking of  {92}  SARS-CoV-2 M$^{\text{pro}}$ inhibitors is of great significance for the future design of anti SARS-CoV-2 drugs.

In this work, for the first time, we predict the binding affinities of these {92} M$^{\text{pro}}$-inhibitor  complexes by reformulate  mathematics-deep learning  (MathDL) models, which have been the top competitor in D3R Grand Challenges, a worldwide competition series in computer aided drug design in the past three years  \cite{nguyen2019mathdl}. We generate reliable poses for 87  M$^{\text{pro}}$ inhibitors with binding affinities but without complex structures. Together with 32 other complexes, we compose a set of 119 M$^{\text{pro}}$-inhibitor complexes, which is paired with  17,382 protein-ligand complexes in PDBbind 2019 general set. These datasets are utilized to construct 11 MathDL models in single-task and multitask settings \cite{nguyen2019mathdl}. One of these 11 MathDL models has been validated by using the PDBbind v2016 core set benchmark, achieving the top performance over all exiting scoring functions. The other ten MathDL models have cross-validated on a set of 119 M$^{\text{pro}}$-inhibitor complexes, showing an averaged Pearson’s correlation coefficient of 0.75.


In a nutshell, the present work provides reliable binding affinity predictions and ranking of 92 SARS-CoV-2 inhibitors that have crystal structures.  It also offers data curation and  validated models for exploring potential SARS-CoV-2 M$^{\text{pro}}$ inhibitors. Furthermore, this work explores different possible binding regions on the SARS-CoV-2 main protease and  decode the most favorable molecular fragments for the inhibitor design.

\section{Results and discussions}

\subsection{Results}\label{sec:results}
\begin{table}[!htb]
\caption{Binding affinities of top 10 complexes in SARS-CoV PBD-noBA dataset  predicted by our MathDL. ``Pred. BA'' indicates the predicted binding free energy in kcal/mol and ``Pred. IC$_{50}$'' is the corresponding IC$_{50}$ in $\mu$M unit.
}
\centering
\begin{tabular}{cccccc}
\toprule
PDBID & Pred. BA & Pred. IC$_{50}$ & PDBID & Pred. BA & Pred. IC$_{50}$\\\midrule
7bqy  & -9.03   &  0.24 & 5rfn & -8.12 & 1.10 \\
5ren  & -8.66   &  0.45 & 5rfh & -8.10 & 1.14 \\
5rfr  & -8.65   &  0.45 & 6w63 & -8.09 & 1.16 \\
5rg1  & -8.57   &  0.52 & 5rer & -7.99 & 1.39 \\
5ret  & -8.29   &  0.82 & 5rfp & -7.95 & 1.48 \\
\bottomrule
\end{tabular}
\label{tab:top10_PDB}
\end{table}

This section is devoted to the utilization of our MathDL models developed in Section \ref{sec:val} to predict the binding affinities and their ranking of SARS-CoV-2 inhibitors that do not have reported experimental affinities. To reduce the role of 3D pose prediction errors in our model, we use the SARS-CoV-2 inhibitors with X-ray structures available in the PDB for our study. We manually search these ligands on the PDB and arrive at the SARS-CoV PDB-noBA dataset consisting of 92 complexes (see Table \ref{tab:datasets}). In this experiment, we develop a MathDL model optimized from PDBbind v2016 core set (see Section \ref{sec:val_coreset}),  five MathDL-ALL and five MathDL-MT models obtained from 5-fold study on the SARS-CoV BA set (see Section \ref{sec:5fold_SARS}). The final predicted binding affinity is the consensus of these 11 models. The top ten inhibitors indicated by our models are shown in Table \ref{tab:top10_PDB}.

The top potent SARS-CoV-2 inhibitor found by our MathDL models is Michael acceptor inhibitor N3 in complex 7bqy. Designed by Yang and his colleagues \cite{yang2005design}, N3  is found viral activities against different coronavirus M$^{\rm pro}$ such as SARS-CoV and MERS-CoV  \cite{yang2005design, wang2016structure}. Specifically, the dissociation constant $K_i$ of N3 was found to be 9.0 $\mu$M against SARS-CoV \cite{yang2005design}. Our MathDL reveals that N3 still inhibits SARS-CoV-2 main protease with an even better affinity at 0.24 $\mu$M. This finding is consistent with   the literature work \cite{jin2020structure} showing that N3 is a potent inhibitor of COVID-19 virus M$^{\rm pro}$. It is worth pointing out N3 and some other molecules in the SARS-CoV PBD-noBA dataset are covalent inhibitors (see discussion in Section \ref{sec:interaction}), however our models only predict the non-covalent binding affinity which is measured before the enzyme deactivation.

Except for complex 6w63  deposited by Mesecar \cite{ghosh2020drug}, the rest of structures reported in Table \ref{tab:top10_PDB} are from PanDDA \cite{PanDDA}. The binding affinities of those PanDDA ligands varying from -7.95 kcal/mol to -8.66 kcal/mol indicate that these fragments can be utilized as starting points when designing more potent SARS-CoV-2 M$^{\rm pro}$ inhibitors. On the other hand, some fragments may not be useful to inhibit the SARS-CoV-2 M$^{\rm pro}$ functions such as U0S (PDBID: 5rgj), T5Y (PDID: 5rf4), and T0V (PDBID: 5re8) with predicted affinity being -3.49 kcal/mol, -4.35 kcal/mol, and -4.55 kcal/mol, respectively. Another factor that can affect their inhibitor abilities is the binding region where its residues and conformation do not bolster the binding mechanism. The predicted binding affinities of all 92 complexes in SARS-CoV PBD-noBA dataset from various MathDL models are presented in Table S8 in Supporting Information. In this table, we also supply the synthetic accessibility score (SAS), partition coefficient $\log P$, and solubility $\log S$ for each small molecule. While the SAS and $\log P$ are obtained via RDkit  \cite{landrum2006rdkit}, the $
\log S$ values are calculated by Alog PS 2.1 \cite{tetko2005virtual}.

\subsection{Discussion}
\subsubsection{Binding site analysis}
\begin{figure}[!ht]
\centering
\includegraphics[scale=0.22]{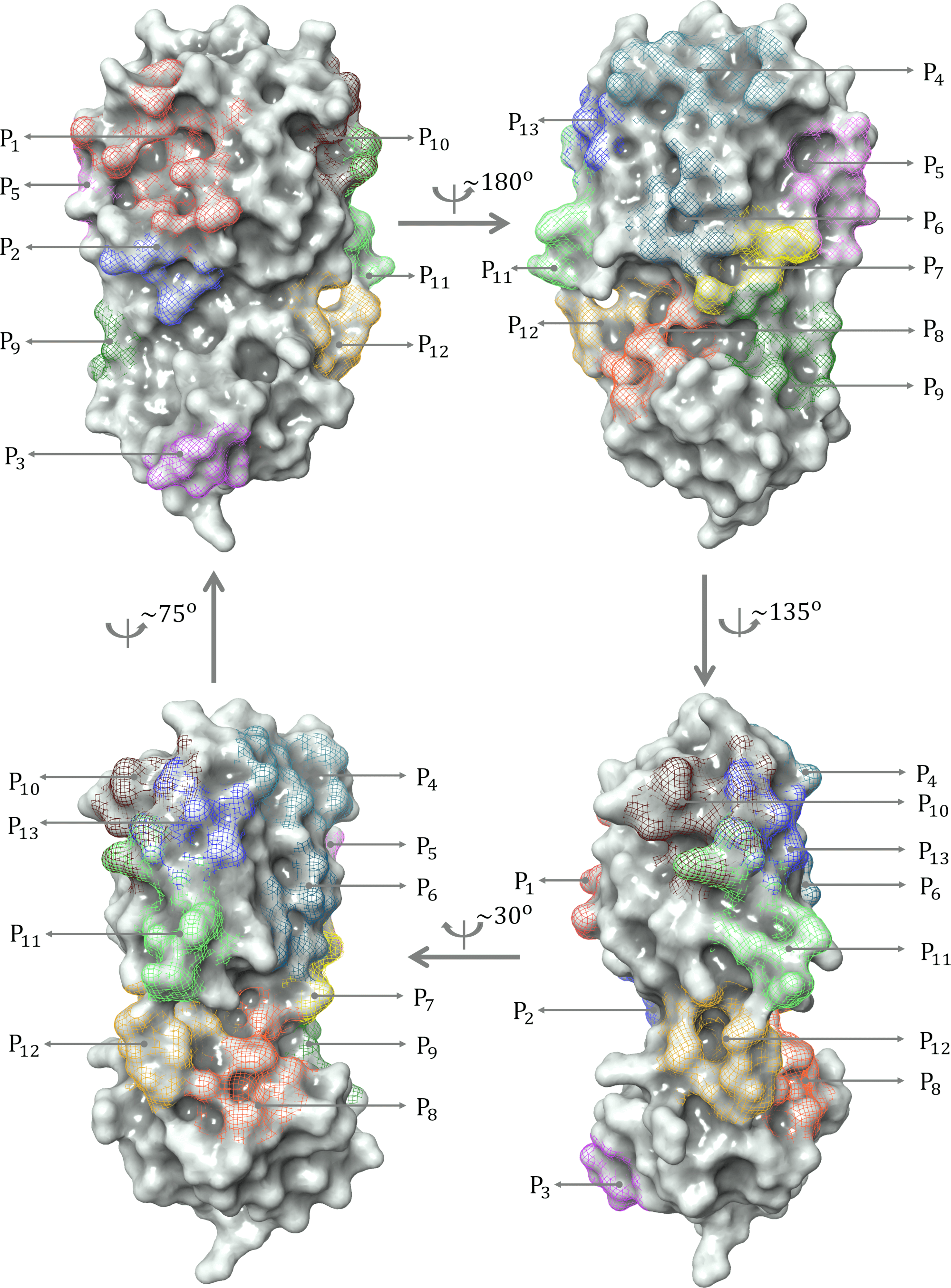}
\caption{All binding site pockets observed from 92 inhibitors in SARS-CoV PDB-noBA set.}
\label{fig:Binding_Sites}
\end{figure}
\begin{figure}[!ht]
\centering
\includegraphics[width=1\textwidth]{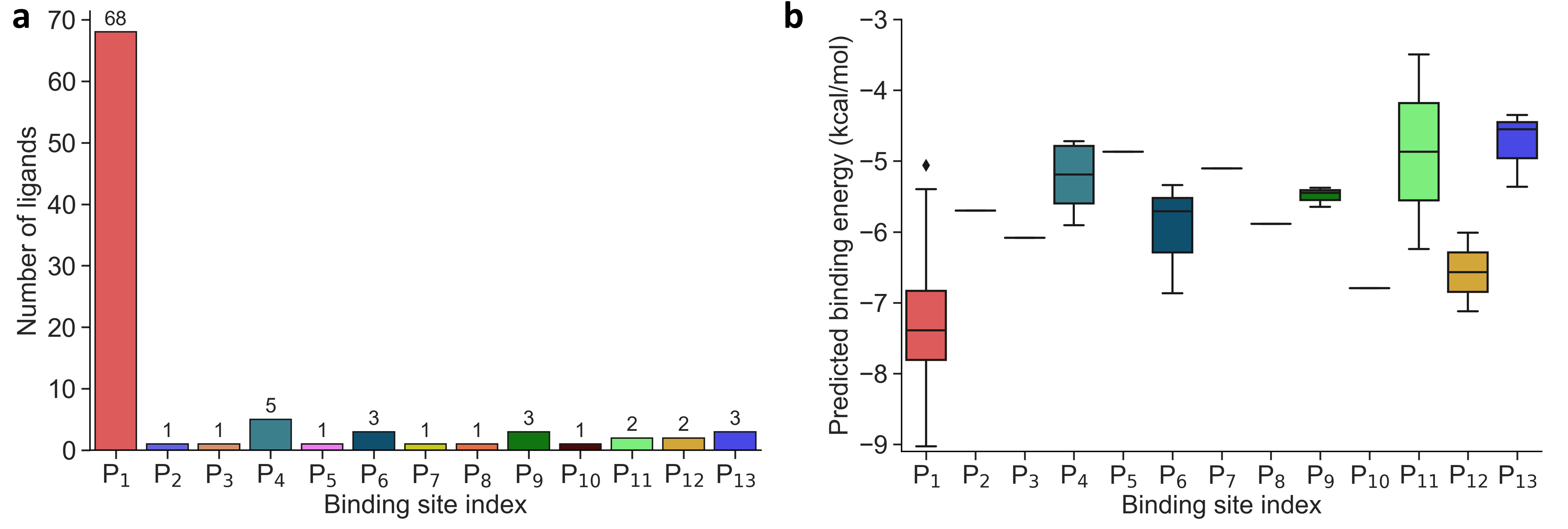}
\caption{a) Distribution of 92 ligands across 11 distinct binding sites; b) Box plot of predicted binding energies (kcal/mol) of all inhibitors in each binding site.}
\label{fig:Binding_Sites_distribution}
\end{figure}

Based on the crystal structure information of 92 complexes in SARS-CoV PDB-noBA set, we have identified 13 distinct binding site regions of the SARS-CoV-2 main protease as illustrated in Figure \ref{fig:Binding_Sites}. Those binding pockets are denoted by ${\rm P}_i, i=1,2,\dots,13$. Figure \ref{fig:Binding_Sites_distribution}a reveals that binding pocket P$_1$ is the most common binding region of the SARS-CoV-2 main protease, which attracts around 73.9\% of ligands in the SARS-CoV PDB-noBA data set of 92 complexes. This finding is no surprise since the binding pocket P$_1$ shares similar active sites to its predecessor, i.e. SARS-CoV M$^{\rm pro}$. Specifically, P$_1$ encompasses His141 and Cys145 catalytic dyad which are imperative to the substrate-binding mechanism \cite{yang2005design}. In additions, the substrate-binding residues Tyr161 and His163 \cite{chang2007reversible} are covered in P$_1$. Binding pockets P$_2$, P$_3$, P$_5$, P$_7$, P$_8$, and P$_{10}$ are the least favor sites consisting of only one ligand. The rest of the binding pockets involve no more than 5 ligands. To study the correlation of the binding regions to the binding free energy, we present the box plot in Figure \ref{fig:Binding_Sites_distribution}b to illustrate the energy values through their quartiles.

The prevailing binding pocket P$_1$ is the best region on the SARS-CoV-2 M$^{\rm pro}$ for inhibitor design with the median binding energy being -7.39 kcal/mol. N3 is the best inhibitor candidate for the binding site P$_1$ with predicted affinity found to be -9.03 kcal/mol. Other binding regions such as P$_{10}$, and P$_{12}$ are less common but show their adequate effects on the binding mechanism with their best energy binding affinities calculated at -6.79 kcal/mol and -7.12 kcal/mol, respectively. These potential binding sites can guide drug combination to inhibit coronavirus M$^{\rm pro}$ effectively.

\subsubsection{Interaction analysis}\label{sec:interaction}
\begin{figure}[!ht]
\centering
\includegraphics[width=1\textwidth]{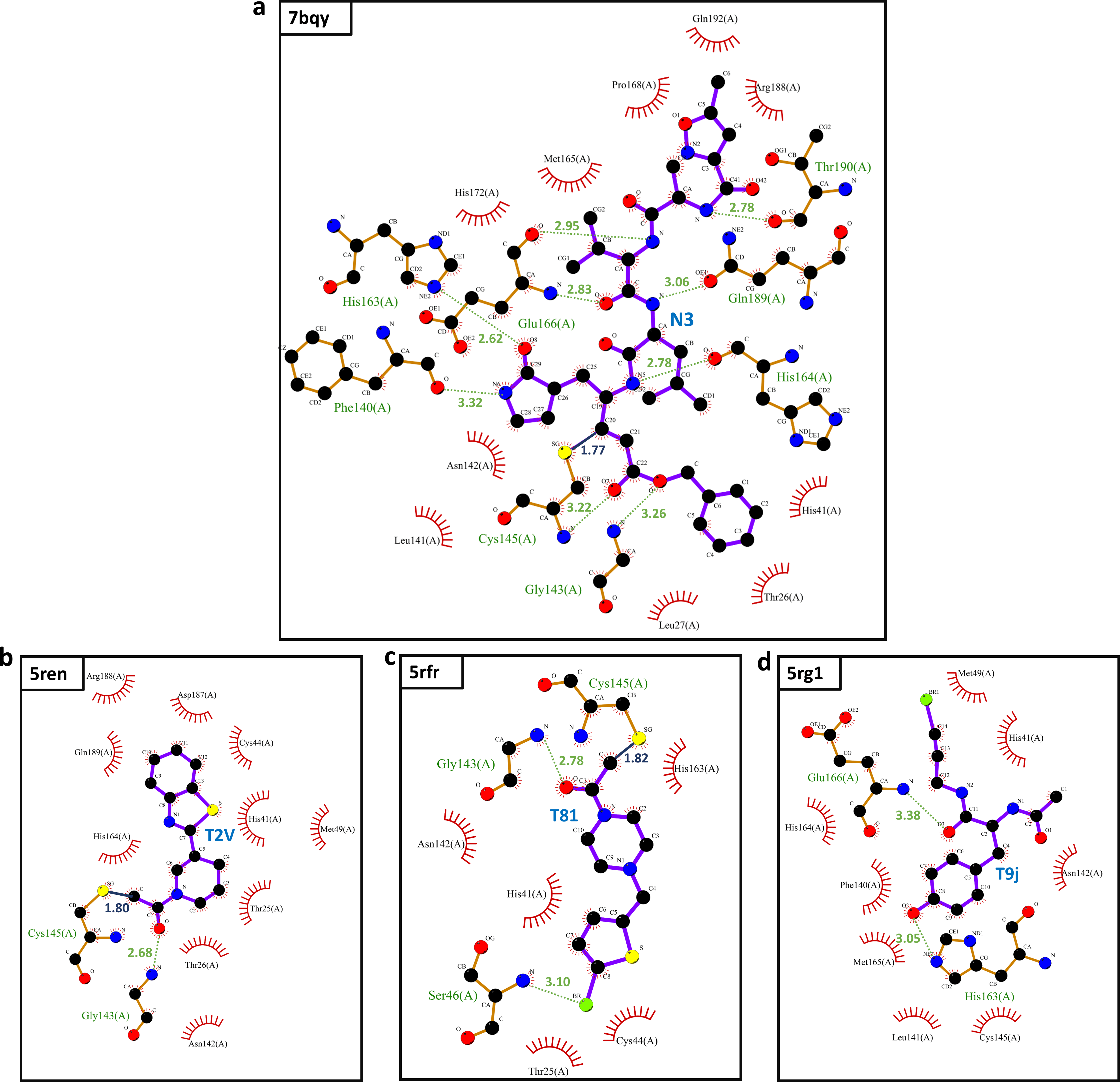}
\caption{The interactions between the top 4 inhibitors in the SARS-CoV PBD-noBA dataset and SARS-CoV-2 M$^{\rm pro}$. Inhibitors are shown in the purple color. Hydrogen bonds are marked in dashed green lines, and covalent bonds are depicted in solid blue lines. All interactions are shown with the distance information in~\AA.}
\label{fig:top4}
\end{figure}

\begin{table}[!htb]
\caption{Interaction analysis in the binding pockets of top 4 complexes in term of binding affinity predicted by our MathDL models.
}
\centering
\begin{tabular}{ccll}
\toprule
PDBID & Ligand ID & Hydrogen bond & Covalent bond\\\midrule
7bqy  & N3   &  Thr190, Gln189, Glu166, Phe140, His164, Gly143, His163, Cys145 & Cys145 \\
5ren  & T2V   &  Gly143 & Cys145\\
5rfr  & T81  &   Gly143, Ser46 & Cys145 \\
5rg1  & T9J   & Glu166, His163 \\
\bottomrule
\end{tabular}
\label{tab:top4_hbonding}
\end{table}

By looking further into the interactions between the top inhibitors and the main protease, we have found that N3 forms the most of hydrogen bonds with 9 interactions and 1 covalent bond to the nearby residues as listed in Table \ref{tab:top4_hbonding} and depicted in Figure \ref{fig:top4}a. All of the hydrogen bonds and covalent bonds found in other complexes in the top 4, namely 5ren, 5rfr, and 5rg1,  are associated with N3's interactions, which justifies the most potent binding of N3 in the main protease complex and confirms the robustness of our MathDL models. We notice that two inhibitors T2V and T81 share one common hydrogen bond to Gly143 and one common covalent bond between their Carbons to $\gamma$-Sulfur of Gly143 in M$^{\rm pro}$ (see Table \ref{tab:top4_hbonding} and Figures \ref{fig:top4}b, \ref{fig:top4}c). Compared to T2V, T81 has one more hydrogen bond but it is a weak one between Brom's T81 and donor from Ser46 at a distance of 3.10~\AA. Therefore, T2V and T81 have very similar predicted binding energies at -8.66 kcal/mol and -8.65 kcal/mol, respectively.
This examination manifests how well our models preserve and capture the physical and chemical properties described in intermolecular bonding interactions. Furthermore, the ligand T9J that binds to M$^{\rm pro}$ in complex 5rg1 with a slightly worse binding energy at -8.57 kcal/mol forms different hydrogen bonds in comparison to two previously mentioned inhibitors (see Table  \ref{tab:top4_hbonding}). Since our models only concern the non-covalent binding affinity, the lack of covalent bond in 5rg1's interactions does not downgrade its binding strength. With two relatively large hydrogen bonding distances (O2-His163: 3.05~\AA, O3-Glu166: 3.38~\AA~(see Figure \ref{fig:top4}d)), the binding affinity of 5rg1 is still comparable to the top inhibitors indicating the important roles in acquiring the hydrogen bonds to these residues in the main protease's binding process.

In the top 10 inhibitors as listed in Table \ref{tab:top10_PDB}, there are two non-covalent inhibitors, namely 5rg1 and 6w63. The rest belongs to the class of targeted covalent inhibitors (TCI) in which the Michael acceptor inhibitor interacts with the protein residues, i.e., cysteine, to form a covalent complex successfully neutralizing target's function. However, the major disadvantage of TCIs is the association with the high toxicity risks \cite{park2011managing}. TCIs' strong covalent bond can dramatically and irreversibly modify the unintended protein targets in the human body. As a result, the top covalent inhibitors in SARS-CoV PBD-noBA dataset may have little chance to become approved market drugs in comparison to their counterparts such as T9j in 5rg1 and X77 in 6w63.

\begin{figure}[!ht]
\centering
\includegraphics[width=0.8\textwidth]{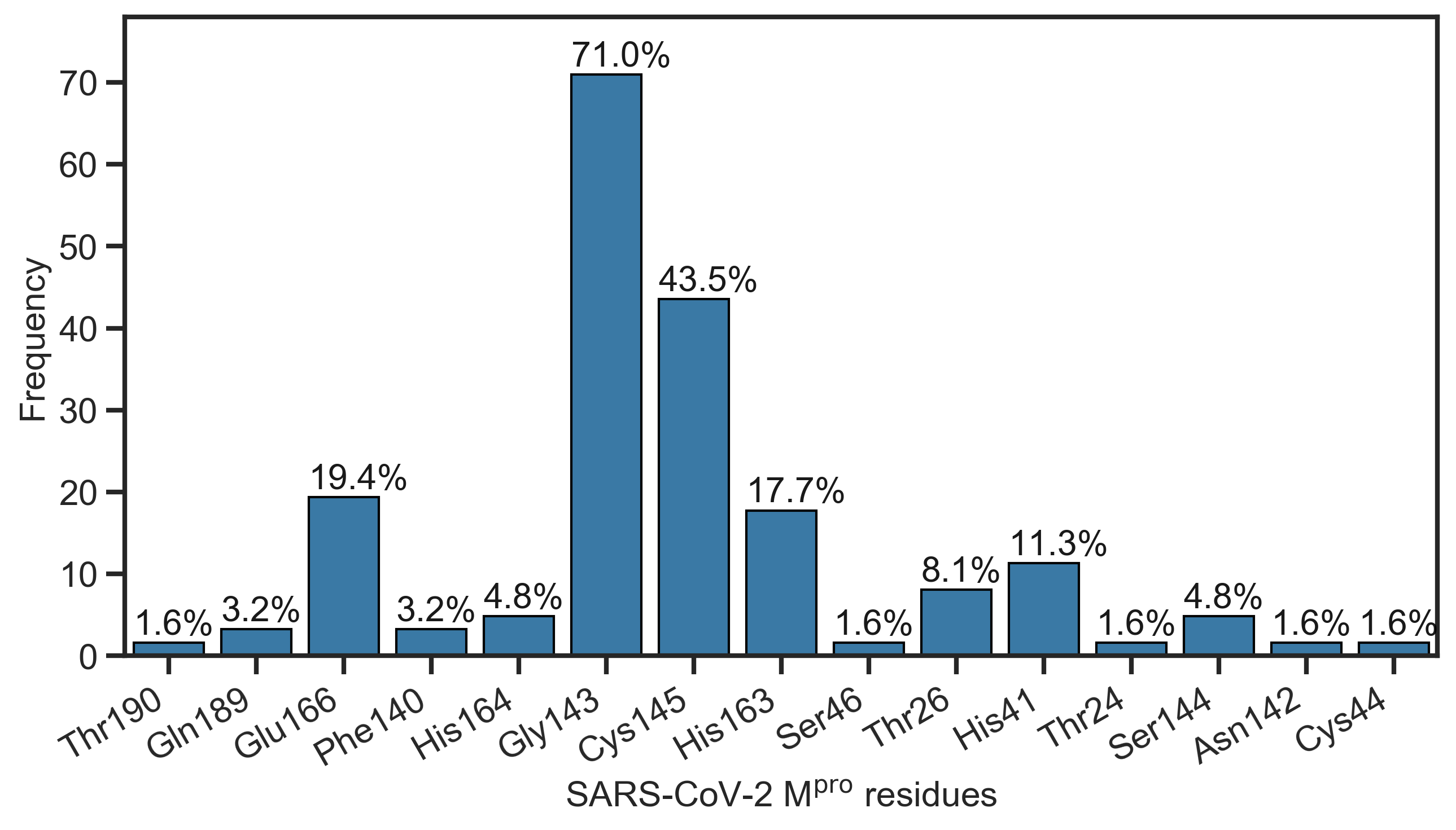}
\caption{Popularity of amino acids in the binding site P$_1$ constituting the hydrogen bonds with ligands.}
\label{fig:hb_frequency}
\end{figure}

\begin{figure}[!ht]
\centering
\includegraphics[width=0.8\textwidth]{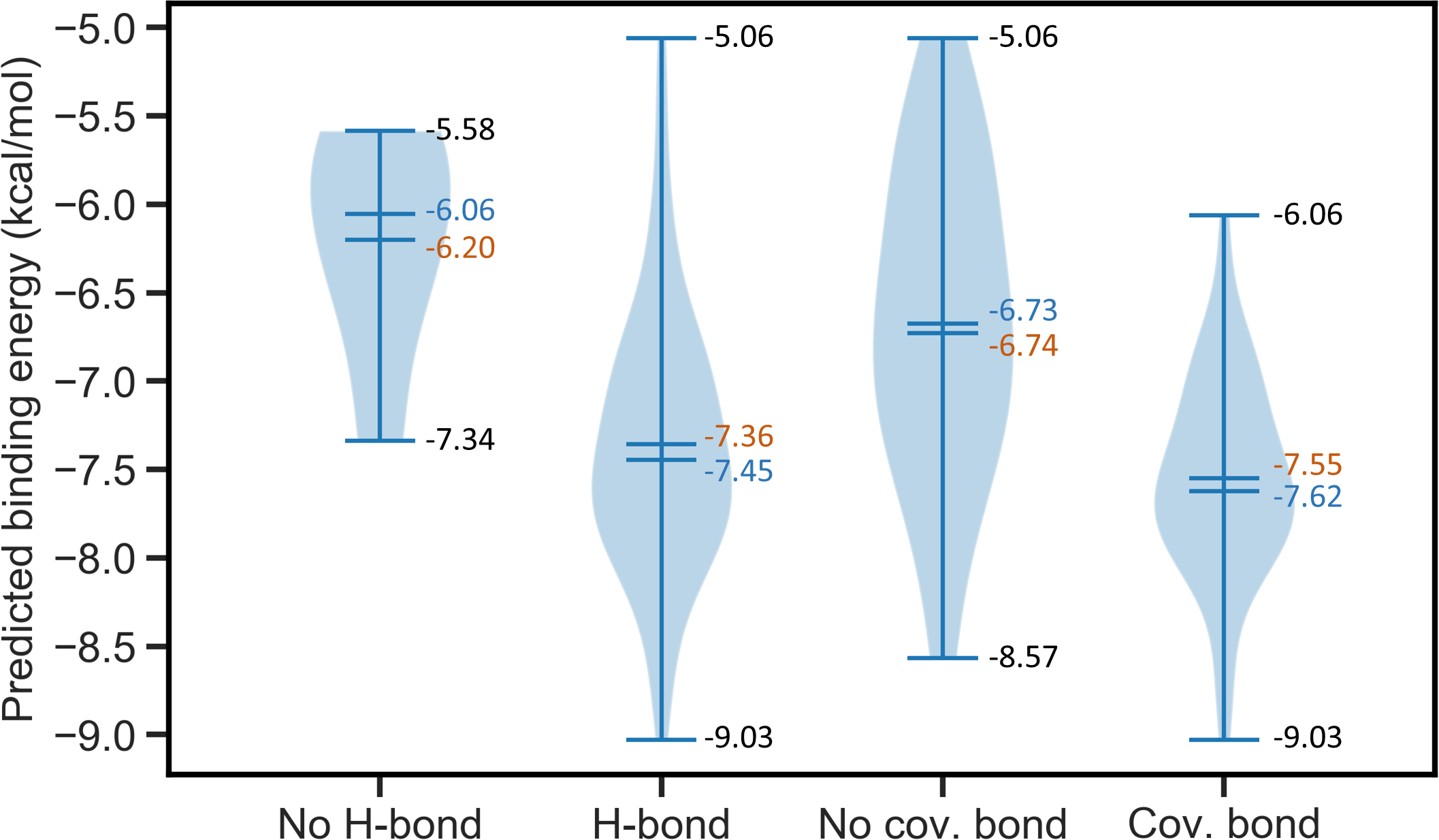}
\caption{Violin plot of the predicted binding energies for 68 inhibitors binding to the binding site P$_1$ classified into 4 categories, namely no H-bond (no hydrogen bond), H-bond (at least one hydrogen bond), no cov. bond (no covalent bond), cov. bond (at least one covalent bond). The mean is in the orange color, the median is in the blue color, and the minimal and the maximal values are both in the black color.}
\label{fig:bond_energy}
\end{figure}
Due to the popularity of the binding site P$_1$ among 92 interested inhibitors, we mainly analyze the interaction network around the residues in that region. Out of 68 molecules binding to P$_1$, there are 62 inhibitors forming at least one hydrogen bond to the nearby amino acid in the SARS-CoV-2 main protease. We have identified 15 different residues in the binding pocket P$_1$ composing hydrogen bonds to these small molecules. Figure \ref{fig:hb_frequency} illustrates the frequency of these 15 residues across 62 inhibitors.  Based on Figure \ref{fig:hb_frequency}, Gly143 residue is the most attractive site to form the hydrogen bond. It appears in 71\% of 62  intermolecular bonding interactions, followed by Cys145  residue with a frequency of 43.5\%. In contrast,  Glu166 residue occupies 19.4\%, while His163 residue occurs in 17.7\%  . It is worth noting when these molecules form a hydrogen bond with  Cys145, they also constitute another hydrogen bond with  Gly143. In all cases, both these residues share the same acceptor. Besides the hydrogen bond network, 45 ligands in the SARS-CoV PDB-noBA dataset form a covalent bond to $\gamma$-Sulfur of Cys145. All the top 3 inhibitors are equipped with that covalent bond, whereas only 2 in the top 10 ligands do not have it (see Table S8 in Supporting Information).

Furthermore, we are interested in the binding energy distribution associated with the interaction network. Figure \ref{fig:bond_energy} depicts the violin plot of that distribution across four categories, namely no H-bond (no hydrogen bond), H-bond (at least one hydrogen bond), no cov. bond (no covalent bond), and cov. bond (at least one covalent bond). Hydrogen bond interactions that are expected to play an important role in the binding mechanism were well captured in our MathDL models. Specifically, while the average energy of inhibitors having none hydrogen bond is -6.2 kcal/mol, the average energy of ones with hydrogen bond is as low as -7.36 kcal/mol.

It is noted that our MathDLs only measure the non-covalent binding affinity. The covalent bond appearing at the final covalent complex is not properly accounted for in our framework. Therefore, it is expected that our models sometimes overestimate the covalent-bond inhibitors over the non-covalent-bond candidates. Figure \ref{fig:bond_energy} reveals molecules in the group of covalent bonds generally are predicted with lower binding energy with an average being -7.55 kcal/mol in comparison to -6.74 kcal/mol averagely measured on ones without covalent bonds.

\subsubsection{Fragment analysis}

\begin{figure}[!ht]
\centering
\includegraphics[width=0.8\textwidth]{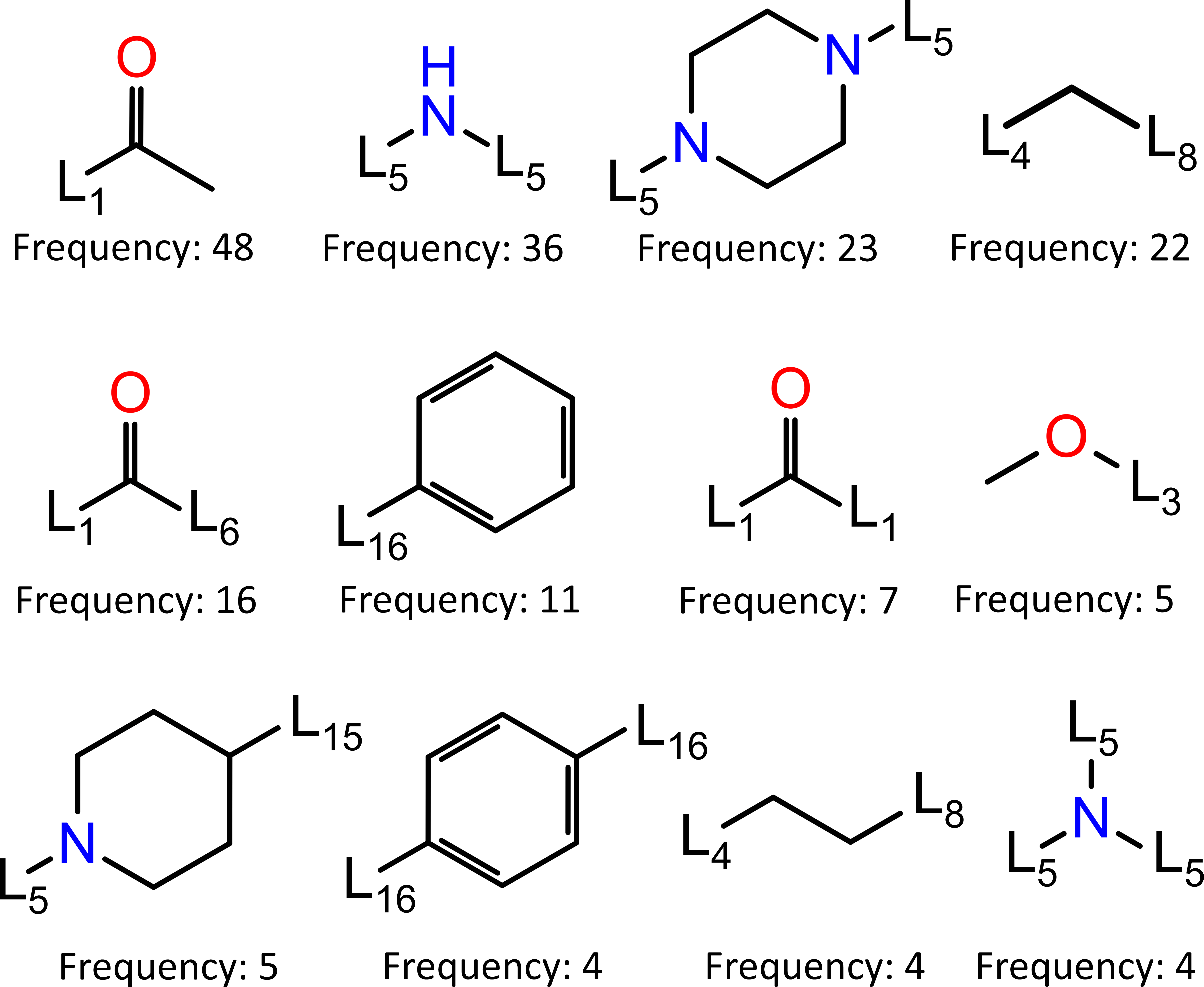}
\caption{Fragment frequencies based on BRICS decomposition of 68 inhibitors of binding site pocket P$_1$. L$_i$ is the link atom of a certain type described in \cite{degen2008art}.}
\label{fig:fragment_distribution}
\end{figure}

To design the lead molecules, it is of importance to have promising fragments from existing inhibitors against the drug targets. Therefore, in the present work, we study all the fragments decomposed from 68 inhibitors attached to the binding site P$_1$. To carry out this task, we utilize BRICS algorithm \cite{degen2008art} via RDkit \cite{landrum2006rdkit}. In BRICS model, there are 16 chemical environments indicated by linkers denoted by L$_1$, L$_2$,\dots,L$_{16}$. The BRICS decomposition gives raise to a total of 126 unique fragments, which are all presented in Table S9 in Supporting Information. Figure \ref{fig:fragment_distribution} illustrates top 12 common fragments in terms of their frequencies. Noting that the top fragment, L$_1$-C(C)=O, often constitutes a hydrogen bond with Gly143 and in many cases forms a covalent bond with Cys145.

\section{Materials and methods}

\subsection{Datasets}\label{sec:datasets}

\begin{table}[!htb]
\caption{A summary of our selected data sets}
\centering
{\renewcommand{\arraystretch}{2}
\begin{tabular}{>{\centering\arraybackslash}m{3.8cm} >{\centering\arraybackslash}m{1.5cm} >{\centering\arraybackslash}m{8cm} >{\centering\arraybackslash}m{2.5cm} }
\toprule
Data name & Data size & Descriptions & References \\\midrule
PDBbind v2019 & 17,382 & Partial PDBbind general set v2019 & \cite{su2018comparative}\\
PDBbind  v2016 core set & 290 &   PDBbind v2016 core set & \cite{su2018comparative} \\
SARS-CoV PDB & 136 & Inhibitors of SARS-CoV/SARS-CoV-2 M$^{\text{pro}}$ having X-ray crystal structures & \cite{Berman:2000,zhang2020crystal, su2020discovery} \\
SARS-CoV PDB-BA & 32 & Inhibitors of SARS-CoV/SARS-CoV-2 M$^{\text{pro}}$ having X-ray crystal structures and experimental binding affinities &\cite{Berman:2000,Berman:2000,su2018comparative,zhang2020crystal,su2020discovery}\\
SARS-CoV PDB-noBA & 92 & Inhibitors of SARS-CoV-2 M$^{\text{pro}}$ having X-ray crystal structures but lacking of experimental binding affinities & \cite{Berman:2000,Berman:2000,su2018comparative,zhang2020crystal,su2020discovery}\\
SARS-CoV 2D & 87 & Inhibitors of SARS-CoV/SARS-CoV-2 M$^{\text{pro}}$ having only 2D structures  & \cite{davies2015chembl,zhang2020crystal, bacha2008development,su2020discovery}\\
SARS-CoV BA & 119 & Inhibitors of SARS-CoV/SARS-CoV-2 M$^{\text{pro}}$ having experimental binding affinities & \cite{Berman:2000,Berman:2000,su2018comparative,su2018comparative,zhang2020crystal,su2020discovery}\\
\bottomrule
\end{tabular}
}
\label{tab:datasets}
\end{table}
Our deep learning-based scoring function, MathDL, was trained on public databases including PDBbind \cite{su2018comparative} and ChemBL \cite{davies2015chembl}. The PDBbind sets contain all complexes with crystal structures deposited in the  PDB  with the binding affinities not limited to Kd, Ki, and IC$_{50}$ reported in the literature. In this work, we employ the PDBbind v2019, the latest version of its generation. The v2019 version of the PDBbind consists of 17,679 protein-ligand complexes. However, the data preprocessing of the MathDL \cite{ZXCang:2018a} only retains 17,382 complexes.

ChemBL is another manually curated database of bioactive molecules. Currently, ChemBL contains more than 2 million compounds in the SMILES string format. Excluding 30 main protease inhibitors in PDBbind data, we have found other 277 small molecules on ChemBL with reported Kd/IC$_{50}$. Additionally, we have found 4 other SARS-CoV main protease inhibitors in \cite{bacha2008development}, and presently 3 SARS-CoV-2 main protease inhibitors from \cite{zhang2020crystal}. In total, there are 314 ligands bound to SARS-CoV/SARS-CoV-2 main protease having the experimental binding affinities; among them, there are 32 crystal structures. For compounds without the crystal structures, MathPose \cite{nguyen2019mathdl} will be utilized to generate their 3D conformations. The predicted 3D coordinates of these structures are presented in the SDF format and available in Supporting Information. Currently, there are roughly 92 ligands forming crystal complexes with SARS-CoV-2 main protease on PDB without the report of the experimental inhibitor activities. Most of them are deposited by the PanDDA analysis group.

To serve model validation purposes, we classify the selected data into five different groups as listed in Table \ref{tab:datasets}. Specifically, PDBbind v2019 is the biggest set in this compilation with its PDBIDs and experimental binding affinities listed in Table S1 in Supporting Information. PDBbind v2016 core set  is a subset of PDBbind v2019 and is formed by 290 complexes representing all protein classes in the refined set of PDBbind v2016 \cite{su2018comparative,nguyen2019agl}. The PDBIDs of all complexes in the  PDBbing v2016 core set  are provided in Table S2. We also collect all M$^{\text{pro}}$ complexes of SARS-CoV/SARS-CoV-2 on the PDB, denoted by SARS-CoV PDB, which results in a total of 136 structures (see Table S3). Among them, there are 32 ligands with the report of experimental binding affinities denoted by SARS-CoV PDB-BA (see Table S4). Furthermore, we are interested in the set of SARS-CoV-2 M$^{\text{pro}}$ complexes in the aforementioned  SARS-CoV PDB set but their affinities are not presented or undisclosed. We call this set SARS-CoV PDB-noBA with PDBIDs listed in Table S5. To enrich our training data targeting SARS-CoV/SARS-CoV-2 main protease inhibitors, we gather some inhibitors reported on the literature \cite{davies2015chembl, bacha2008development}. For those compounds with only 2D information, we limit ourselves to ones having the similarity score based on the path-based fingerprint FP2 no lower than 0.6 to at least one inhibitor in the SARS-CoV PDB set. As a result, we arrive at a set of 87 structures named SARS-CoV 2D (see Table S6). Combining SARS-CoV PDB-BA and SARS-CoV 2D data sets, we finalize a reliable database focusing on SARS-CoV/SARS-CoV-2 main protease inhibitors. Table S7 in Supporting Information presents the PDBIDs as well as the experimental binding energies of these ligands.

\subsection{Methods}
\subsubsection{MathDL}\label{sec:MathDL}
\begin{figure}[ht]
\centering
\includegraphics[width=1\textwidth]{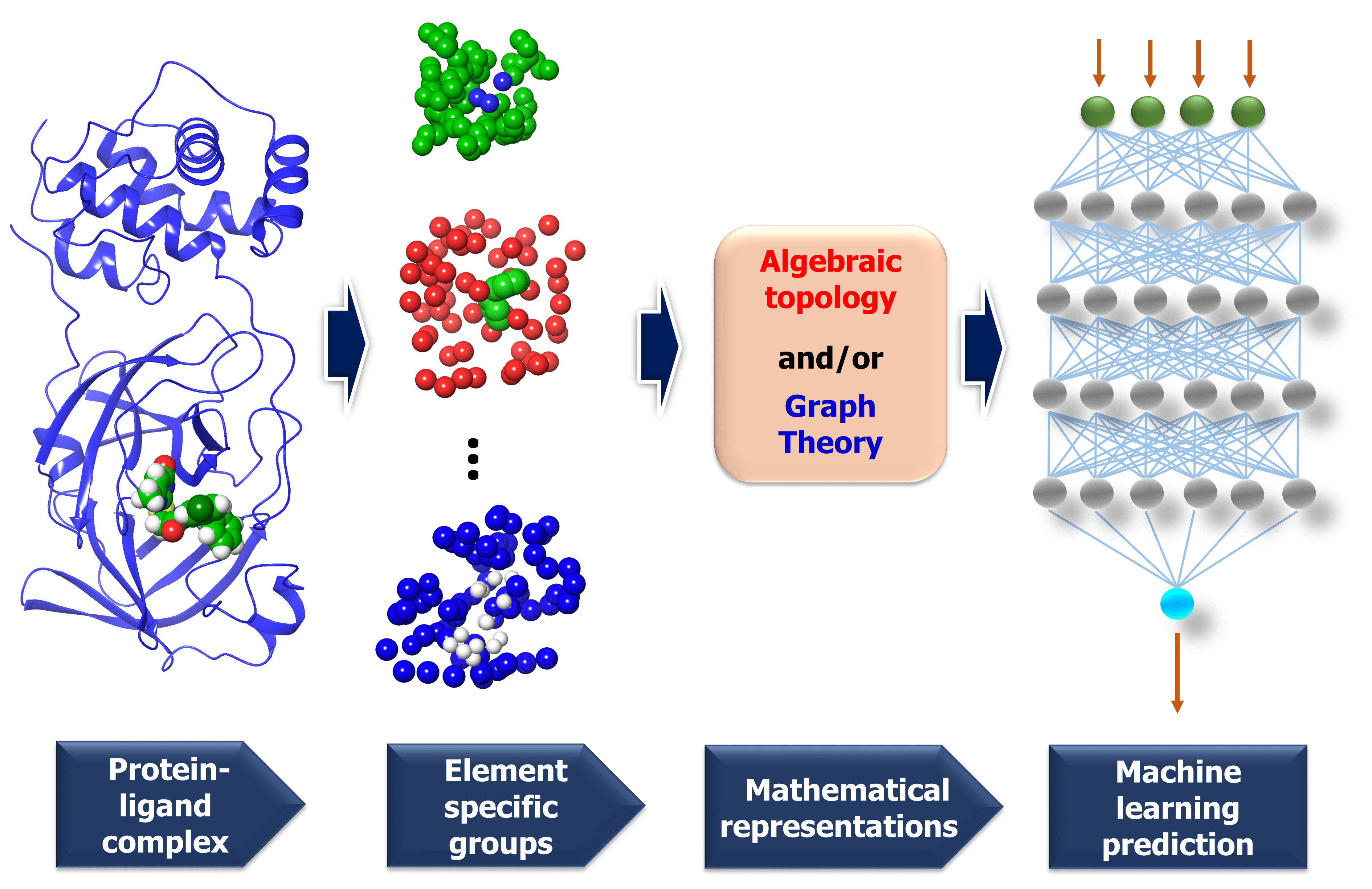}
\caption{A framework of MathDL energy prediction model which integrates advanced mathematical representations with sophisticated CNN architectures.}
\label{fig:MathDeep}
\end{figure}

The MathDL models developed in this work are reformulated from our early model bearing the same name. MathDL was designed for the prediction of various druggable properties of 3D molecules \cite{nguyen2019mathdl}. In the past three years, MathDL has been proved to be the top competitor in D3R Grand Challenges (\url{https://drugdesigndata.org/about/grand-challenge}), a worldwide competition in computer-aided drug design. 
I the present work, we have, for the first time,  develop a multitask MathDL (MathDL-MT) to handle the M$^{pro}$ inhibitor dataset. We have also extended our earlier MathDL by including all different datasets (MathDL-All). \autoref{fig:MathDeep} depicts the framework of the MathDL in which the element-specific algebraic topological representations are integrated with the convolutional neural network (CNN) aiming to predict varied druggable properties such as toxicity, binding affinities, etc.

\paragraph{Algebraic topology-based representations.}
Algebraic topology studies the topological spaces with the use of abstract algebra, which can dramatically simplify the geometric complexity. Persistent homology (PH) is one of the algebraic topology approaches which has the capacity to track the multiscale topological information over different scales along with filtration by characterizing independent components, rings, and higher dimensional voids in space \cite{carlsson2009topology}. In this section, we will briefly review the algebraic topology-based representations. Additionally, since we are dealing with the protein-ligand system, therefore, the biological considerations will take into account as well.

\justifying
{\bf Simplex.}
The $q$-simplex denoted as $\sigma_q$ is the convex hull of $q + 1$ affinely independent points in $\mathbb{R}^n \ (n \ge k)$. For example, the $0,1,2, \text{and }3$-simplex is considered as a vertex, an edge, a triangle, and a tetrahedron, respectively. We call the convex hull of each non-empty subset of $q+1$ points the face of $\sigma_q$, and each points are also called the vertices.

\justifying
{\bf Simplicial complex.}
A set of simplices is a simplicial complex denote $K$ which satisfies that every face of a simplex $\sigma_q \in K$ is also in $K$ and the non-empty intersection of any two simplices in $K$ is the common face for both.

\justifying
{\bf Chain complex.}
A formal sum of $q$-simplices in simplicial complex $K$ with $\mathbb{Z}_2$ coefficients is a $q$-chain is a. A set of all $q$-chains of the simplicial complex $K$ equipped with an algebraic field (typically $\mathbb{Z}_2$) is called a chain group and denoted as $C_q(K)$. The boundary operator is defined by $\partial_q: C_q(K) \to C_{q-1}(K)$ to relate the chain groups. More specifically, we denote $\sigma_q = [v_0, v_1, \cdots, v_q]$ for the $q$-simplex spanned by its vertices, and then the boundary operator can be represented as:
\begin{equation}
    \partial_q \sigma_q = \sum_{i=0}^{q}(-1)^i\sigma^{i}_{q-1}.
\end{equation}
Here, $\sigma^{i}_{q-1} = [v_0, \cdots, \hat{v_i},\cdots,v_q]$ is the $(q-1)$-simplex with $v_i$ being omitted. The sequence of chain groups connected by boundary operators is called the chain complex and expressed as:
\[
\cdots \stackrel{\partial_{q+2}}\longrightarrow C_{q+1}(K) \stackrel{\partial_{q+1}}\longrightarrow C_{q}(K) \stackrel{\partial_{q}}\longrightarrow C_{q-1}(K)\stackrel{\partial_{q-1}} \longrightarrow \cdots
\]
The $q$-cycle group $Z_q(K)$ and the $q$-boundary group $B_q(K)$ are defined as $Z_q(K) = ker(\partial_q) = \{c\in C_q(K) \ | \ \partial_q c = \emptyset \}$ and $B_q(K) = im(\partial_{q+1}) = \{\partial_{q+1}c \ | \ c \in C_{q+1}(K) \}$. The $q$-th homology group is the quotient group $H_q(K) = Z_q(K)/B_q(K)$. Moreover, the rank of $q$-th homology group can be computed as $\text{rank}H_q(K) = \text{rank}Z_q(K) - \text{rank}B_q(K)$, which is denoted as the $q$-th Betti number $\beta_q$. To be notice that the $q$-th Betti number count the number of $q$-dimensional holes that can not be continuously deformed to each other.

\justifying
{\bf Persistent Homology.}
A filtration of a simplicial complex $K$ is a nested sequence of subcomplexes of $K$ such that $\emptyset = K_0 \subseteq K_1 \subseteq K_2 \cdots \subseteq K_m = K $. Then the $p$-persistent $q$th homology group of $K_t$ is defined as:
\begin{equation}
    H_q^p(K_t) = Z_q(K_t)/(B_q(K_{t+p}) \cap Z_q(K_t)).
\end{equation}
Here the rank of $H_q^p(K_t)$ counts the number of $q$-dimensional holes in $K_t$ that are still alive in in $K_{t+p}$, which is called the $p$-persistent $q$th Betti number. The persistent homology not only records the topological information at a specific configuration, but also tracks the changes along with the filtration parameters. More specifically, the topological changes will be preserved in the persistent barcodes. In MathDL, we make use of the persistent homology barcodes by dividing them into bins and calculating the birth, death, and persistence incidents in each bin to enrich our algebraic topological representations.

\paragraph{Element specific considerations.}
The protein-ligand complex is structural and also biological. The persistent homology provides a theoretical approach to encode high-dimensional spatial data of protein-ligand complexes into algebraic topological representations. In this section, we address the biological considerations for biomolecular complexity. There are many kinds of interactions that exist in the protein-ligand complex, such as electrostatics, hydrogen bonds, and hydrophobic effects. Although persistent homology can capture the interactions among the nearest neighbors, the long-range interactions will be hindered. This difficulty can be avoided via the deployment of the element-specific attention \cite{ZXCang:2018a}. There are 4 commonly atom types in protein, namely {C, N, O, S}, and there are 11 commonly atom types in ligand, including {C, N, O, S , P, F, Cl, Br, I, H, B}. We include Boron in the ligand atom type consideration since  it appears in more than 200 small compounds in our training data. The general framework of MathDL is depicted in Figure \ref{fig:MathDeep} under exemplified steps. For the details of feature descriptions as well as the deep learning architecture, interested readers are referred to our previous work \cite{ZXCang:2018a}.

\subsubsection{MathPose} \label{sec:MathPose}

MathPose, a 3D pose predictor which converts SMILES strings into 3D poses with references of target
molecules, was the top performer in D3R Grand Challenge 4  (GC4) in predicting the poses of 24 beta-secretase 1 (BACE) binders \cite{nguyen2019mathdl}. For one SMILES string, around 1000 3D conformations can be generated by various docking software tools such as GOLD \cite{G-Score}, Autodock Vina \cite{Trott:2010AutoDock}, and GLIDE \cite{friesner2004glide}. Moreover, a selected set of known complexes is re-docked by three aforementioned docking software packages to generate at least 100 decoy complexes per input ligand used in the machine learning training set. The machine learning labels will be the calculated root mean squared deviations (RMSDs) between the decoy and native structures for the training data of the pose selection task. Furthermore, MathDL models will be set up and applied to select the top-ranked pose for the given ligand. Besides the GC4 challenge, our models have outperformed state-of-the-art scoring functions at the docking power challenge on CASF-2007 and CASF-2013 benchmarks \cite{nguyen2019agl}. Those established results attest to the credibility of our MathPose on the 3D structure prediction of small molecules.

\subsection{Validations}\label{sec:val}

\subsubsection{PDBbing v2016 core set benchmark}\label{sec:val_coreset}

\begin{figure}[!htb]
    \centering
    \includegraphics[width=1\textwidth]{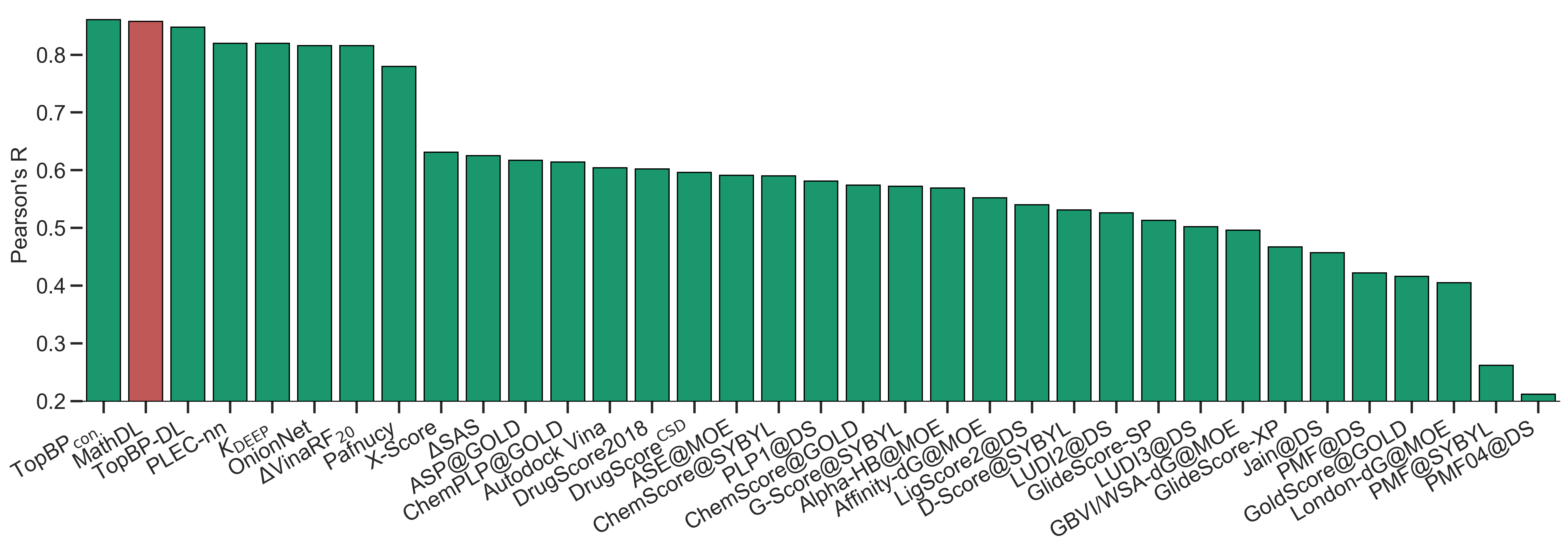}
    \caption{The Pearson correlation coefficient of various scoring functions on PDBbing  v2016 core set benchmark. Our MathDL is in red. The performances of other models that are in green are taken from Refs. \cite{su2018comparative,    wojcikowski2018development,  nguyen2019dg, zheng2019onionnet,ZXCang:2018a}. TopBP$_{\rm con.}$, the consensus model in our published work \cite{ZXCang:2018a}, attains the highest $R_p$ at 0.861. The current MathDL is followed with the second highest $R_p$ at 0.858 and ${\rm RMSE}=1.56$ kcal/mol. The third place in the list is another TopBP model, TopBP-DL, solely based on the deep learning architectures and its reported $R_p$ is 0.848  \cite{ZXCang:2018a}. It is noted that all of the machine learning based scoring functions in this comparison were trained on the PDBbind v2016 refined set of 3767 complexes except for our MathDL. Explicitly, MathDL is trained on a much larger training set consisting of 17,211 complexes picked out from the PDBbind v2019 set and SARS-CoV BA set.}
    \label{fig:coresetv2016-benchmark}
\end{figure}

In this validation task, we will testify our model against 290 complexes in the PDBbing v2016 core set. This is a prevalent test set to assert the scoring ability of a binding affinity prediction model and has attracted lots of research groups to devote the effort to improve the Pearson's correlation coefficient ($R_p$) and Kendall's tau ($\tau$) on this core set performance \cite{su2018comparative,nguyen2020review,jimenez2018k}. In the current work, we merge the PDBbind v2019, SARS-CoV PDB-BA, and SARS-CoV 2D sets but removing the duplicates and excluding the PDBbing v2016 core set complexes to attain a training set of 17211 complexes. MathDL with the architecture described in Section \ref{sec:MathDL} is trained on those complexes. The resulting model is utilized to predict the binding affinity of 290 structures in the PDBbing v2016 core set.

With the purpose of exploring the most optimal model for this benchmark, MathDL is trained for 1000 epochs. Then, we pick the epoch based on the root-mean-squared error (RMSE) of the PDBbing v2016 core set   prediction. We have found that MathDL achieves the smallest RMSE in this experiment at 140 epochs. Specifically RMSE, $R_p$, and $\tau$ metrics on the  v2016 core set are 1.56 kcal/mol, 0.858, and 0.671, respectively. Meanwhile, the training accuracy is 0.387 kcal/mol in terms of RMSE and its Pearson's correlation coefficient is $R_p=0.994$. These performances reveal that our MathDL converges very fast and with only 140 epochs and maintains a good balance between training and testing accuracies. This is a state-of-the-art performance since our MathDL is ranked in the second place in comparison to 33 other scoring functions (see Figure \ref{fig:coresetv2016-benchmark}). It is noted that the top model is TopBP$_{\rm con.}$ published in our previous work \cite{ZXCang:2018a} with $R_p=0.861$. TopBP$_{\rm con.}$ is the consensus of gradient boosted tree and deep learning-based models. If only the deep learning framework is considered, the performance of TopBP (denoted by TopBP-DL) on the core set of PDBbind v2016 is $R_p=0.848$.

It is worth mentioning that except for our MathDL, all machine learning-based scoring functions listed in Figure \ref{fig:coresetv2016-benchmark} were trained on the  PDBbind v2016 refined set of 3767 complexes. As mentioned above, the current MathDL is compiled on a much larger training set comprised of 17211 complexes selected from PDBbind v2019 and  SARS-CoV BA data. Even the present MathDL has not outperformed its predecessor, i.e., TopBP$_{\rm con.}$, MathDL is still a preference model since it is trained on a diverse data set covering various protein families and different binding energy ranges. As a result, it is expected to deliver more reliable predictions on the SARS-CoV-2 inhibitor, especially when this main protease family is not included in the training data of previous TopDL models. The resulting MathDL model is labeled as MathDL-Core2016 and is utilized to predict affinities of complexes in SARS-CoV PDB-noBA  in Section \ref{sec:results}.

\subsubsection{5 fold cross-validation on SARS-CoV BA set}\label{sec:5fold_SARS}
\begin{table}[!htb]
\caption{5-fold Performances of MathDL-All and MathDL-MT on SARS-CoV BA set}
\centering
\begin{tabular}{lcccccc}
\toprule
 & \multicolumn{3}{c}{MathDL-ALL} & \multicolumn{3}{c}{MathDL-MT}\\
\cmidrule(lr){2-4} \cmidrule(lr){5-7}
 & $R_p$ & $\tau$ & RMSE & $R_p$ & $\tau$ & RMSE \\\midrule
 Fold 1 (Train) & 0.996 & 0.950 & 0.303 &  0.994 & 0.937 & 0.254 \\
 Fold 1  (Test) & 0.830 & 0.616 & 0.675 &  0.835 & 0.623 & 0.671\\
 \midrule
 Fold 2 (Train) & 0.993 & 0.935 & 0.415 & 0.998 & 9.967 & 0.187 \\
 Fold 2 (Test) & 0.657 & 0.490 & 0.755 &  0.612 & 0.417 & 0.766\\
 \midrule
 Fold 3 (Train) & 0.998 & 0.966 & 0.206 & 0.999 & 0.975 & 0.100\\
 Fold 3 (Test)  & 0.660 & 0.526 & 1.041 & 0.692 & 0.548 & 1.009\\
 \midrule
 Fold 4 (Train) & 0.999 & 0.979 & 0.124 & 0.999 & 0.975 & 0.104\\
 Fold 4 (Test) & 0.824 & 0.599 & 0.883  & 0.781 & 0.621 & 0.947\\
 \midrule
 Fold 5 (Train) & 0.994 & 0.936 & 0.368 & 0.993 & 0.926 & 0.311\\
 Fold 5 (Test) & 0.786 & 0.534 & 0.981 & 0.813 & 0.542 & 0.916\\
 \midrule
 Average (Train) & 0.996 & 0.953 & 0.208 & 0.997 & 0.956 & 0.191\\
 Average (Test) &  0.751 & 0.553 & 0.867 & 0.747 & 0.550 & 0.862\\
 \bottomrule
\end{tabular}
\label{tab:5-fold}
\end{table}

In this section, we testify the performance of our MathDL against 119 inhibitors in the SARS-CoV BA set aforementioned in Table \ref{tab:datasets}. Among those ligands, there are 32 X-ray crystal structures and the rest is in 2D SMILES string. We employ MathPose to predict 3D structures of those 2D ligands. To carry out the validation, we randomly split the SARS-CoV BA set into 5 non-overlapped folds. In each fold prediction task, MathDL trains on the partial data of SARS-CoV BA in conjunction with PDBbind v2019 set. This situation results in two different ways of training our MathDL model. The first approach is a traditional MathDL architecture with the training set combining both SARS-CoV BA and PDbbind v2019 complexes. The second model makes use of multi-task learning \cite{KDWu:2018a}. In each epoch, the weights of the MathDL architecture are learned through the information from PDBbind v2019 set, then only the fully connected layers are trainable when learning SARS-CoV BA structures. Finally, we come up with 10 different MathDL models in which the traditional MathDL frameworks are labeled as MathDL-All-$i$ and multi-task MatDL is named MathDL-MT-$i$ with $i$ running from 1 to 5. In each model, after 100 epochs, we start monitoring which epoch that helps our model achieve the smallest RMSE on the test set.

Table \ref{tab:5-fold} reveals that MathDL-All models are well trained with the averaged accuracy RMSE=0.208 kcal/mol,  Pearson's correlation coefficient $R_p$=0.996, and Kendall's tau $\tau=$ 0.953.  Their averaged performances on test data across 5-fold of the SARS-CoV BA set are found to be $R_p$=0.751, $\tau=$ 0.553, and RMSE=0.867 kcal/mol. These results endorse the reliability of these models in the binding affinity prediction of SARS-CoV/SARS-CoV-2 inhibitors. Table \ref{tab:5-fold} also lists the training and testing performances of five multi-task learning models. The averaged training performance of the MathDL-MT model is $R_p$=0.997, $\tau$=0.956 and RMSE=0.191 kcal/mol. The accuracy of the multi-task architecture on the test sets is similar to MathDL-All with $R_p$=0.747, $\tau$=0.55, and RMSE=0.862 kcal/mol. With these promising results, it is encouraging to carry out MathDL models to predict unknown binding affinities of SARS-CoV/SARS-CoV-2 inhibitors.

\section{Conclusion}
 SARS-CoV-2 main protease  (M$^{\text{pro}}$) is the most favorable target for COVID-19 drug discovery due to its conservative nature and low similarity with human genes. Structure and binding affinity of protein-drug complexes are of paramount importance for understanding the molecular mechanism in drug discovery. However, there are only two SARS-CoV-2 M$^{\text{pro}}$ inhibitor structures available with binding affinities, highlighting current challenges in COVID-19 drug discovery.

This work presents the reliable binding affinity prediction and ranking of 92 M$^{\text{pro}}$-inhibitor crystal structures that have no reported experimental binding affinity.  We first curate a set of 314 M$^{\text{pro}}$ inhibitors with binding affinities from public resources,  such as PDBbind,  ChemBL and the scattered literature.  Among these inhibitors,  87 are retained based on their high similarity with available M$^{\text{pro}}$-inhibitor complex structures and  built with three dimensional (3D) poses using our MathPose   \cite{nguyen2019mathdl}.  Together with 32 another SARS-CoV or SARS-CoV-2 M$^{\text{pro}}$-inhibitor complexes, we compose a training set of  119 reliable  SARS-CoV-2 M$^{\text{pro}}$-inhibitor complexes. Our earlier  MathDL models are reformulated to accommodate 119 new complexes and  17,382 complexes from the PDBbind v2019 general set in both single-task and multitask settings, which have never been available before.
The resulting MathDL models are rigorously validated via PDDbind v2016 core set benchmark in which it outperforms   state-of-the-art models  in the literature. Most importantly, our MathDL achieves promising cross-validation accuracies on the SARS-CoV family inhibitors with the averaged Pearson's correlation coefficient as high as 0.75.

Additionally,  the present work unveils that Gly143 of M$^{\text{pro}}$ is the most attractive region to form a hydrogen bond, followed by Cys145, Glu166, and His163. There are 45 inhibitors interacting with SARS-CoV-2  M$^{\text{pro}}$ to form covalent complexes. Those covalent bonds are mostly composed between dicarbon monoxide groups in  inhibitors and $\gamma$-Sulfur on Cys145. There are only two non-covalent complexes in our top 10 ranked, namely 5rg1 and 6w63. To provide a potential resource for lead molecule design, we employ 
the BRICS algorithm to decompose all the inhibitors of the prominent binding site on M$^{\text{pro}}$ and obtain 126 unique fragments.

The predicted binding affinities and their ranking of 92 M$^{\text{pro}}$-inhibitor crystal structures, the bonding analysis, and the fragment decomposition have significantly extended current knowledge and understanding of  SARS-CoV-2 M$^{\text{pro}}$ and inhibitor interactions and thus offered valuable information toward COVID-19 drug discovery.

\section*{Supporting Information}
\begin{itemize}
  \item \url{SupportingTables.xls}: Spreadsheets contain information for all supporting tables from S1 to S8.
  \item \url{FileS1.zip}: 3D structures generated by our MathPose for 87 ligands in SARS-CoV 2D set.
\end{itemize}

\section*{Acknowledgments}
This work was supported in part by NIH grant GM126189, NSF Grants DMS-1721024, DMS-1761320, and IIS1900473, Michigan Economic Development Corporation, Bristol-Myers Squibb, and Pfizer.
The authors thank The IBM TJ Watson Research Center, The COVID-19 High Performance Computing Consortium, and  NVIDIA for computational assistance.


\end{document}